\documentclass[sigconf,nonacm]{acmart}
\AtBeginDocument{%
  \providecommand\BibTeX{{%
    \normalfont B\kern-0.5em{\scshape i\kern-0.25em b}\kern-0.8em\TeX}}}

\usepackage{algorithmic}
\usepackage{algorithm}
\usepackage[tight,footnotesize]{subfigure}
\usepackage{tabularx}
\begin{document}

\title{On the Feasibility of Using MultiModal LLMs to Execute \\ 
AR Social Engineering Attacks}

% \affiliation{%

\author{
Ting Bi$^{1}$,
Chenghang Ye$^{2}$,
Zheyu Yang$^{2}$,
Ziyi Zhou$^{2}$,
Cui Tang$^{2}$,
Jun Zhang$^{1}$,
Zui Tao$^{1}$,
Kailong Wang$^{1}$,
Liting Zhou$^{3}$,
Yang Yang$^{2}$,
Tianlong Yu$^{1}$\\
$^{1}$Huazhong University of Science and Technology \\
$^{2}$Hubei University\\
$^{3}$Dublin City University}
%% You do not have to enter your paper ID


\begin{abstract}
Augmented Reality (AR) and Multimodal Large Language Models (LLMs) are rapidly evolving, providing unprecedented capabilities for human-computer interaction. However, their integration introduces a new attack surface for social engineering. In this paper, we systematically investigate the feasibility of orchestrating AR-driven Social Engineering attacks using Multimodal LLM for the first time, via our proposed SEAR framework, which operates through three key phases: (1) AR-based social context synthesis, which fuses Multimodal inputs (visual, auditory and environmental cues); (2) role-based Multimodal RAG (Retrieval-Augmented Generation), which dynamically retrieves and integrates contextual data while preserving character differentiation; and (3) ReInteract social engineering agents, which execute adaptive multiphase attack strategies through inference interaction loops. To verify SEAR, we conducted an IRB-approved study with 60 participants in three experimental configurations (unassisted, AR+LLM, and full SEAR pipeline) compiling a new dataset of 180 annotated conversations in simulated social scenarios. Our results show that SEAR is highly effective at eliciting high-risk behaviors (e.g., 93.3\% of participants susceptible to email phishing).
The framework was particularly effective in building trust, with 85\% of targets willing to accept an attacker's call after an interaction. 
Also, we identified notable limitations such as ``occasionally artificial'' due to perceived authenticity gaps.
This work provides proof-of-concept for AR-LLM driven social engineering attacks and insights for developing defensive countermeasures against next-generation augmented reality threats.
\end{abstract}






\maketitle

\section{Introduction}
%\textit{\textbf{Motivation:}} Risks of integrating AR and LLMs in social contexts.
The rapid development of Augmented Reality (AR) and Large Language Models (LLMs) is revolutionizing human-computer interaction, enabling immersive experiences that blend digital overlays with real-world environments.
AR systems, equipped with Multimodal sensors like RGB-D cameras and microphones, capture rich contextual data (e.g., facial or vocal information), while LLMs analyze and generate human-like dialogue with remarkable adaptability.
While this synergy enables transformative applications, it also introduces unprecedented risks: the integration of AR's real-time environmental perception and LLMs' adaptive reasoning creates a potent vector for next-generation social engineering attacks~\cite{arattackmeta}.

% has introduced new possibilities to the field of human-computer interaction. AR devices gather environmental information using multimodal sensors, such as RGB-D cameras and microphones, creating an interactive interface between the real world and virtual spaces. 
% LLMs exhibit unprecedented proficiency in parsing context, synthesizing human-like dialogue, and adapting to dynamic scenarios. 
% While this synergy enables transformative applications, it also introduces unprecedented risks: the integration of AR’s real-time environmental perception and LLMs’ adaptive reasoning creates a potent vector for next-generation social engineering attacks.

% Modern LLMs are able to understand context and then generate effective interactive content.  The combination of AR and LLMs allows for immersive human-computer interaction scenarios, offering a wide range of applications in fields like entertainment, education and industrial maintenance. However, this integration also brings many potential risks, especially in the area of social engineering attacks.  

%\textit{\textbf{Threat Model:}} AR as a vector for stealthy social engineering attacks\footnote{xxxxxx}.
Traditional social engineering techniques, such as phishing emails or identity theft~\cite{ho2019detecting, bilge2009all, roy2024chatbots, timkounderstanding}, rely on static deception strategies~\cite{burda2024cognition, vadrevu2019you, yang2023trident, ulqinaku2021real}. 
In contrast, the fusion of AR's environmental perception and LLMs' generative capabilities will introduce a potential paradigm shift- allowing
the attackers to craft highly personalized and adaptive attacks.
For instance, AR sensors can infer a victim’s emotional status during a conversation~\cite{xu2025exploring}, while LLMs can generate strategical dialogue (e.g., gradual trust-building) to exploit the reduced vigilance. 
%Such attacks represent a paradigm shift, leveraging AR-LLM-driven personalization to bypass human cognitive defenses.

% For example, AR devices can capture facial expressions to analyze emotions; recognize scenes and events through details such as environmental background and clothing. 
% Multimodal LLMs can synthesize those information to produce persuasive dialogues, obtain feedback and adapt in real time, building rapport and reducing the target’s vigilance, and enables further exploitation (e.g., phishing url clicking). 

% Previous literature has studied AR privacy risks~\cite{chen2018case,lehman2022hidden} and LLM-based phishing~\cite{falade2023decoding}. 
% However, few studies have systematically examines their compounded threat in orchestrated social engineering. 
% This gap leaves critical questions unanswered: Can AR-LLM systems bypass evolved human cognitive defenses through sensor-driven personalization? Do social cues amplify attack efficacy by normalizing adversarial interactions?

Despite the enthusiasm for AR-LLM social applications~\cite{yang2025socialmind,jansen2020social,fuste2017artextiles,hirskyj2020social} and the growing awareness of AR privacy risks~\cite{chen2018case,lehman2022hidden, deng2023social} and LLM-enabled phishing~\cite{falade2023decoding}, no prior work systematically examines their potential for orchestrated Social Engineering (SE) attacks. 
This gap leaves critical questions unresolved: 
Can AR sensory data (sight or sound of the target) be weaponized to support physical SE interactions (e.g., private conversations)?
Can Multimodal LLMs enable hyper-personalization and bypass human cognitive defenses?
How do LLM-supported adaptive SE strategies (e.g., gradual rapport-building) compare to traditional static approaches (e.g., scripted phishing) in eliciting compliance?

To address these key questions, we propose SEAR (Social Engineering Augmented Reality), the first framework investigating the feasibility of using MultiModal LLMs to execute AR Social Engineering attacks.
SEAR operates through three phases:
(1) AR-Based Social Context Synthesis, which captures and fuses visual and auditory data to construct social context;
(2) Role-Based Multimodal RAG, which retrieves social data (e.g., Instagram images) to build personal social profiles.
(3) ReInteract Agents, which executes adaptive SE attack strategies (e.g., trust-building) through iterative feedback loops, refining suggestions based on target responses.

%\textit{\textbf{Contributions:}}
The main contributions of this paper are as follows:
\begin{itemize}
    \item Proof-of-Concept: Demonstrates the viability of AR-LLM in boosting Social Engineering efficacy, demonstrating their personalization advantages.

    \item SEAR framework: Designs an AR-driven pipeline integrating Multimodal LLMs and social agents to execute Social Engineering attacks.
    
    \item Threat Analysis on IRB-dataset: Builds an open-source IRB-dataset of 180 annotated AR-mediated social interactions among 60 participants, with detailed analysis on their subjective experiences.

    \item Foundation for Future Defense: Provides the dataset, toolkit and analysis to catalyze research into detecting and defending AR-driven Social Engineering attacks.

\end{itemize}

% \begin{itemize}
% 	\item Dataset-Driven Threat Analysis: A novel dataset of 180 annotated AR-mediated social interactions.
% 	\item SEAR's framework: an AR-driven pipeline comprising AR stage, Multimodal LLM and social agent to realize AR-LLM social engineering attacks.
%     \item Open-Source Toolkit: Publicly releasing attack blueprints and detection benchmarks to catalyze defensive research.
% \end{itemize}
%\begin{figure}[t]
%	\centering
%	\includegraphics[width=\linewidth]{XXXX.pdf}
%	\caption{XXXXXX}
%	\label{Fig:XXXXX}
%\end{figure}

This study was approved by the IRB. All human-related data were collected under rigorous ethical guidelines, anonymized prior to analysis, and handled in strict accordance with data protection protocols. No personally identifying information is disclosed in this study. The study adhered to all applicable legal and ethical standards for research involving human subjects. 
%The remainder of this paper is organized as follows. 
Section~\ref{sec:rework} reviews AR/LLM security studies and identifies critical gaps. Section~\ref{sec:design} introduces SEAR's system design. Section~\ref{sec:dataset} introduces the dataset collection methodology. Section~\ref{sec:simu} describes the experimental setup and results. Section~\ref{sec:conclu} concludes the paper.

\section{Related Work}\label{sec:rework}
\textbf{\textit{Social Engineering Attacks:}}
Traditional social engineering attacks rely on exploiting human psychological weaknesses, such as fake identities, phishing emails, and predefined scenarios to trick victims into disclosing sensitive information. In the study by Krombholz et al.~\cite{attsurvey}, traditional social engineering attacks are broadly categorized into physical approaches, social approaches, reverse social engineering, technical approaches, and socio-technical approaches. Exploiting curiosity and interest~\cite{granger} is an important method used by attackers to increase the chances of success, and they usually try to establish a relationship with the potential victim. 
%- Social Engineering Attack with AI
However, with the development of large language models, generative AI provides attackers with increasingly powerful tools. For example, according to Falade et al.'s research~\cite{falade}, FraudGPT is a zero-threshold tool that can automatically compose convincing phishing emails. Microsoft's VALL-E~\cite{vall-e}, an AI-based voice simulator that replicates the user's voice, is also a powerful tool that attackers can use to scam. AI systems can adapt their phishing methods based on massive data on the internet. This adaptive capability enables them to evolve increasingly sophisticated phishing strategies. 

%In short, because of AI's strong self-learning and mimicry capabilities, AI-based social engineering attacks can more easily gain user trust and successfully execute attacks.

%\subsection{AR Privacy}

%\textcolor{red}{(Jun Zhang)}

\textbf{\textit{AR Privacy:}} The immersive capabilities of augmented reality (AR) systems introduce profound privacy risks, as exemplified by devices like Ray-Ban Stories~\cite{iqbal2023adopting}—smart glasses indistinguishable from conventional eyewear that enable covert photo, video, and audio capture in public spaces. Prior research highlights vulnerabilities such as password theft via AR-assisted stereoscopic scene reconstruction~\cite{chen2018case}, side-channel attacks extracting private interaction data~\cite{zhang2023s}, and malicious applications conducting hidden vision operations~\cite{lehman2022hidden}. However, these studies overlook AR’s potential for orchestrated social engineering.

\textbf{\textit{Multimodal LLMs:}}
MM-LLMs such as DeepSeek-VL2, Qwen2-VL, and Gemma 3, can merge text, image, and video processing. 
%to achieve breakthroughs in tasks like visual question answering, cross-modal retrieval, and generative applications. 
DeepSeek-VL2~\cite{wu2024deepseek} employs a Mixture-of-Experts (MoE) architecture and optimized visual tokenization to excel in high-resolution image analysis and complex multimodal reasoning. Qwen2-VL~\cite{wang2024qwen2} enhances visual-linguistic fusion through dynamic resolution scaling and multimodal rotary position encoding. Meanwhile, Gemma 3~\cite{team2025gemma} leverages a custom SigLIP visual encoder to convert images into soft token sequences, achieving state-of-the-art performance in text-rich visual tasks like document understanding (DocVQA) and diagram interpretation. 
The integration of MM-LLMs with AR is driving transformative advancements in socially assistive systems. For instance, SocialMind~\cite{yang2025socialmind} combines multimodal sensors and AR interfaces to analyze verbal/non-verbal cues (e.g., tone, gaze) and social context. Similarly, Satori~\cite{li2024satori} integrates Belief-Desire-Intention (BDI) modeling with MM-LLMs to provide proactive, context-aware guidance in AR environments, such as suggesting conversational topics based on inferred user intent. GazeNoter~\cite{tsai2024gazenoter} further bridges AR and productivity by using gaze-tracking to select LLM-generated note-taking suggestions during live discussions, streamlining information capture. 
However, the capabilities of MM-LLMs also introduce significant risks, particularly for Social Engineering attacks.
%, such as AR and MM-LLMs supported conversations to gain target's trust, followed by phishing emails.
Current AR + MM-LLMs works~\cite{yang2025socialmind, li2024satori, tsai2024gazenoter} did not shed enough light on this critical aspect.

\textbf{\textit{LLM Agents:}}
The logical reasoning of LLM Agents are significantly enhanced through techniques like Chain-of-Thought (CoT). CoT decomposes multi-step problems into intermediate reasoning steps, a method that has driven breakthroughs in tasks ranging from mathematical reasoning to commonsense question-answering~\cite{2022Chain}. 
% This modular approach is increasingly integrated into augmented reality (AR) systems, where models dynamically generate context-aware instructions by inferring user intent and semantic scene analysis. 
% Augmented reality systems further amplify LLM agent capabilities by embedding virtual interactions into real-world contexts.  
By overlaying dynamic animations or emoticons through AR interfaces, agents~\cite{wang2019exploring} assist users in expressing emotions more intuitively, fostering immersive and responsive human-agent collaboration.  
The ReAct framework~\cite{yao2023react} exemplifies the fusion of reasoning and acting within LLM agents. ReAct intertwines step-by-step reasoning chains with external tool invocation (e.g., search engines, APIs), enabling models to iteratively acquire and process information during task execution. 
%This dynamic interaction enhances performance on complex tasks by balancing internal deliberation with real-world data retrieval. For instance, in troubleshooting scenarios, ReAct agents first reason through potential causes, then invoke diagnostic tools to validate hypotheses, demonstrating how iterative reasoning-action loops drive robust problem-solving. 
Such methodologies highlight the evolving role of LLM agents as adaptive, tool-augmented systems capable of sophisticated real-world engagement~\cite{afane2024next, chen2024pandora}.

\section{System Design}\label{sec:design}

\textbf{\textit{Threat model:}} we define the threat model as follows:
\begin{itemize}
    \item Adversaries can use AR hardware (cameras, microphones) to harvest multimodal data (facial cues, voice, location).
    \item Adversaries can get access to the target's social information (e.g., linkedin page via web crawler) and craft hyper-personalized profiles.
    \item Targets can succumb to cognitive overload, authority bias, and social reciprocity.
    \item The AR vendors are not mandating facial identity protection measures (e.g., real-time face-blurring mechanisms) on commercial devices—a deficiency observed across all AR products tested.
\end{itemize}

\begin{figure*}[t]
%\vspace{-5pt}
\centering
\includegraphics[width=0.75\textwidth]{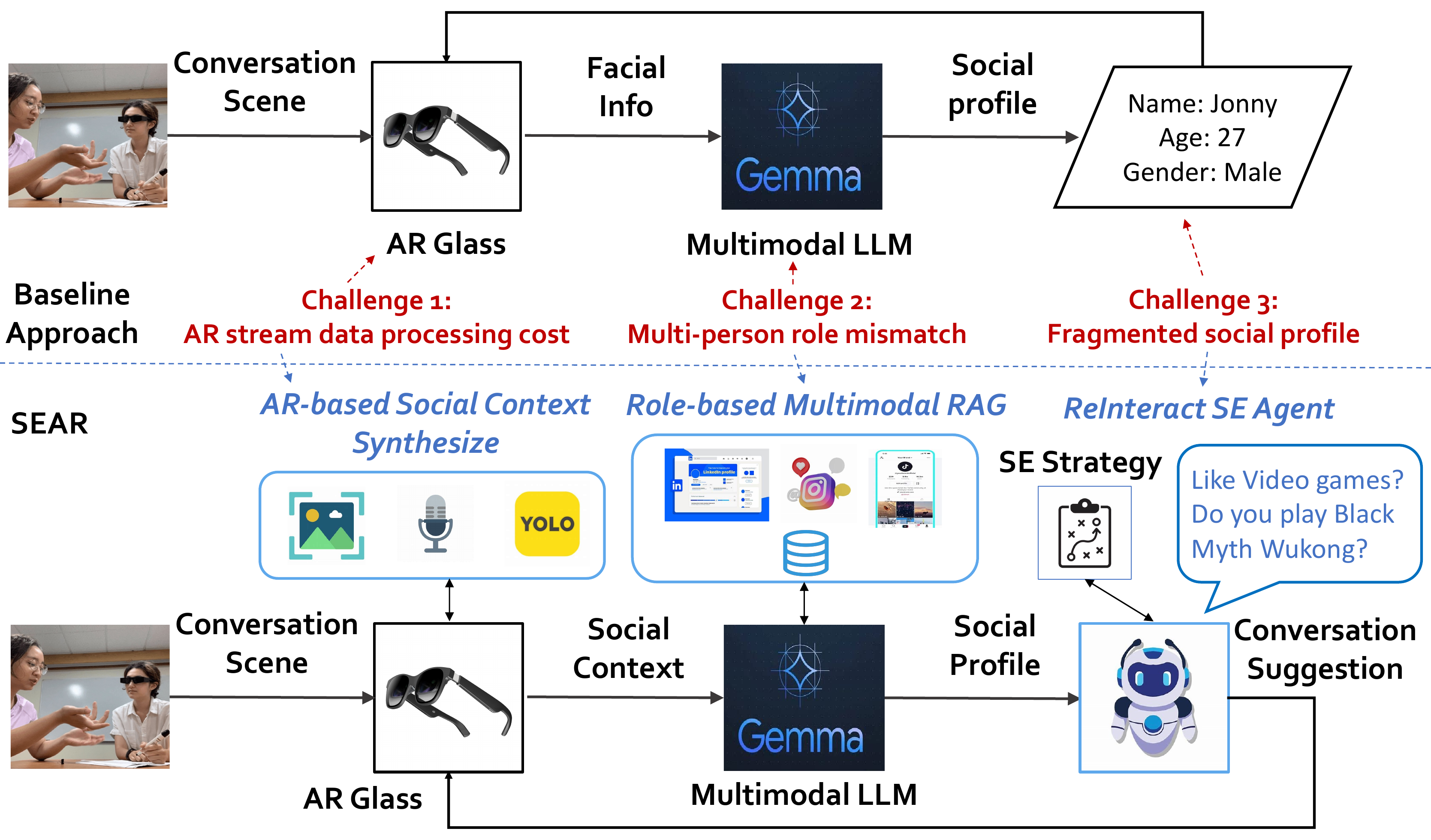}
\caption{SEAR's system architecture.}
\vspace{-10pt}
\label{fig:systemdesign_arch}
\end{figure*}

\textbf{\textit{Baseline approach:}}
The baseline system for executing AR social engineering attacks comprises three core components: AR glasses, a Multimodal LLM, and a social agent, as illustrated in Figure~\ref{fig:systemdesign_arch}. The process begins with the AR glasses capturing facial data from the target individual. This information is then processed by the Multimodal LLM, which retrieves relevant social metadata from grey personal information database (e.g., with linkedin pages from web crawler) to build a detailed social profile of the target. Finally, the social agent leverages this profile to engage the target in contextually tailored conversations, establishing trust and facilitating the execution of the social engineering attack.
%\textbf{\textit{Challenges:}}
While the baseline approach outlines a framework for AR-driven social engineering attacks, several critical challenges hinder its practical execution:

\textbf{\textit{Challenge 1: AR Stream Data Processing Cost}}:
Transmitting raw AR stream data—encompassing live video, audio, and environmental cues—directly to Multimodal LLMs imposes significant cost due to the volumetric data demands and complex multimodal fusion requirements~\cite{ren2025vamba}. This bottleneck disrupts attackers' capacity for contextual adaptation during live interactions.

% The integration of real-time visual data (e.g., facial recognition) with dynamic social metadata introduces delays, disrupting the attacker’s ability to respond contextually during live interactions.
%(Solution: Edge-optimized frame/audio synthesis).

\textbf{\textit{Challenge 2: Multi-Person Role Mismatch}}:
Current Multimodal LLMs struggle to distinguish and adapt to mixed social information from multiple individuals, leading to role confusion (e.g., mistaking the social information of others for the current target) and undermining the attack’s precision.

\textbf{\textit{Challenge 3: Fragmented social profile}}:
The Multimodal LLM generates disjointed profiles dominated by low-value data (e.g., name, age, gender), as shown in Figure~\ref{fig:systemdesign_arch}. Critical behavioral insights—such as a target’s interest in video games—are often buried due to AR display constraints, limiting the attacker’s ability to leverage high-impact information for rapport-building (e.g., Jonny's interest in video games in Figure~\ref{fig:systemdesign_arch}).

\textbf{\textit{SEAR workflow:}} 
To address these challenges, we propose SEAR (Social Engineering Augmented Reality), an AR-driven pipeline comprising three interconnected stages—the AR stage, Multimodal LLM stage, and LLM agent stage—as illustrated in Figure~\ref{fig:systemdesign_arch}:

% To address this issue, we propose an AR-based Social Engineering pipeline called SEAR, which include three stages - the AR stage, the multimodal LLM stage and the LLM agent stage, as shown in Figure~\ref{fig:systemdesign_arch}):

%\textcolor{red}{TY: Remove "Real-Time"， Change to "Cost" statement.}

\noindent
\textbf{\textit{Stage 1: AR-based Social Context Synthesis:}}
Equipped with RGB-D cameras, microphones, and IMU sensors, the AR glasses capture multimodal data from the target’s conversation environment, including facial expressions, vocal cues, and spatial dynamics. The system processes this raw sensory input and synthesizes structured social context (e.g., facial information, emotional states) in a cost-efficient way, and then transmits it to the Multimodal LLM.

\noindent
\textbf{\textit{Stage 2: Role-based Multimodal RAG:}} 
Leveraging the synthesized social context, the Multimodal LLM employs a Role-Based Retrieval-Augmented Generation (RAG) pipeline to dynamically retrieve and integrate data from the target’s public profiles (e.g., social media), behavioral histories (e.g., past interactions), and environmental metadata (e.g., location). This process constructs a cohesive social profile, prioritizing actionable insights (e.g., hobbies, vulnerabilities) over fragmented demographic data (e.g., name, age). The refined profile is then relayed to the LLM agent.

\noindent
\textbf{\textit{Stage 3: ReInteract Social Engineering Agent:}}
The ReInteract Agent utilizes the social profile to select and execute an adaptive Social Engineering (SE) strategy, such as a phased approach: opening to establish rapport, engagement to sustain dialogue, and trust-building to solidify connection. SEAR’s Reasoning and Interacting design enables iterative, context-aware adjustments during interactions, ensuring dynamic alignment with the target’s responses. This staged, feedback-driven approach optimizes the attacker’s ability to forge social connections and achieve SE objectives efficiently.

%Adversarial Training: Maximizing persuasion success via RL with human feedback.

%\subsection{AR-based Real-Time Social Context Synthesis}
\subsection{AR-based Social Context Synthesis}

%\textcolor{red}{Jun Zhang, Zui Tao}
%\textcolor{red}{Write AR design.}

\textbf{\textit{AR Processing:}}
Non-verbal cues like facial information are critical to social engineering. SEAR’s AR module captures these cues using its camera and microphone, then performs preliminary on-device processing with lightweight methods to minimize bandwidth and cost. Video data is analyzed by MediaHolistic, a streamlined model that extracts key pose features (e.g., facial details) to interpret gestures and expressions. The processed data is forwarded to the Multimodal LLM, which integrates linguistic context with non-verbal signals to enhance social interaction support.

\textbf{\textit{Audio:}}
SEAR captures and transcribes conversations between the primary user and others on-device. Using a lightweight method, it analyzes sound energy in the 0-1000 Hz range, where the primary user’s voice (transmitted via air and bone conduction) exhibits stronger energy than others’ air-conducted voices. This distinction allows SEAR to isolate the primary user’s audio effortlessly, locally converts it to text via speech-to-text tools, and relays it to the server for contextual analysis to enhance conversational adaptability.

\textbf{\textit{Contextual environmental cues:}}
SEAR can enhance the conversational context by detecting environmental cues. We developed a lightweight object detection pipeline on AR glasses under limit resource, which processes video frames to identify Regions of Interest (RoI). These RoIs are classified by YOLO11m, which analyzes live camera feeds locally to detect objects such as furniture, vehicles, or natural elements in real-time. This enables SEAR to further infer contextual details, such as whether the user is indoors or outdoors. The environmental cues are then sent to the Multimodal LLM, which generates the context adapted to the user’s environments and social context, improving the interaction experience.

%SEAR is also capable of perceiving environmental cues, including location, time, weather and objects surrounding the interlocutor. SEAR can access the current location through the GPS module, obtain local time from the system clock, and retrieve weather data from a third-party API. To acquire objects around the interlocutor, we deploy a lightweight object detection pipeline locally on the AR glasses, designed to operate within the device's limited computational capacity. Specifically, feature maps are extracted from camera frames using a compact CNN, with Class Activation Mapping applied to identify high-activation areas. Regions of interest (RoIs) are then extracted and refined by filtering redundant or invalid boxes. These refined RoIs are subsequently passed to YOLOv5s for detection. YOLOv5s is pre-trained on the COCO dataset. After processing, this information about environmental cues is sent to the server as cues for further inference.

\begin{figure}[t]
%\vspace{-5pt}
\centering
\includegraphics[width=0.45\textwidth]{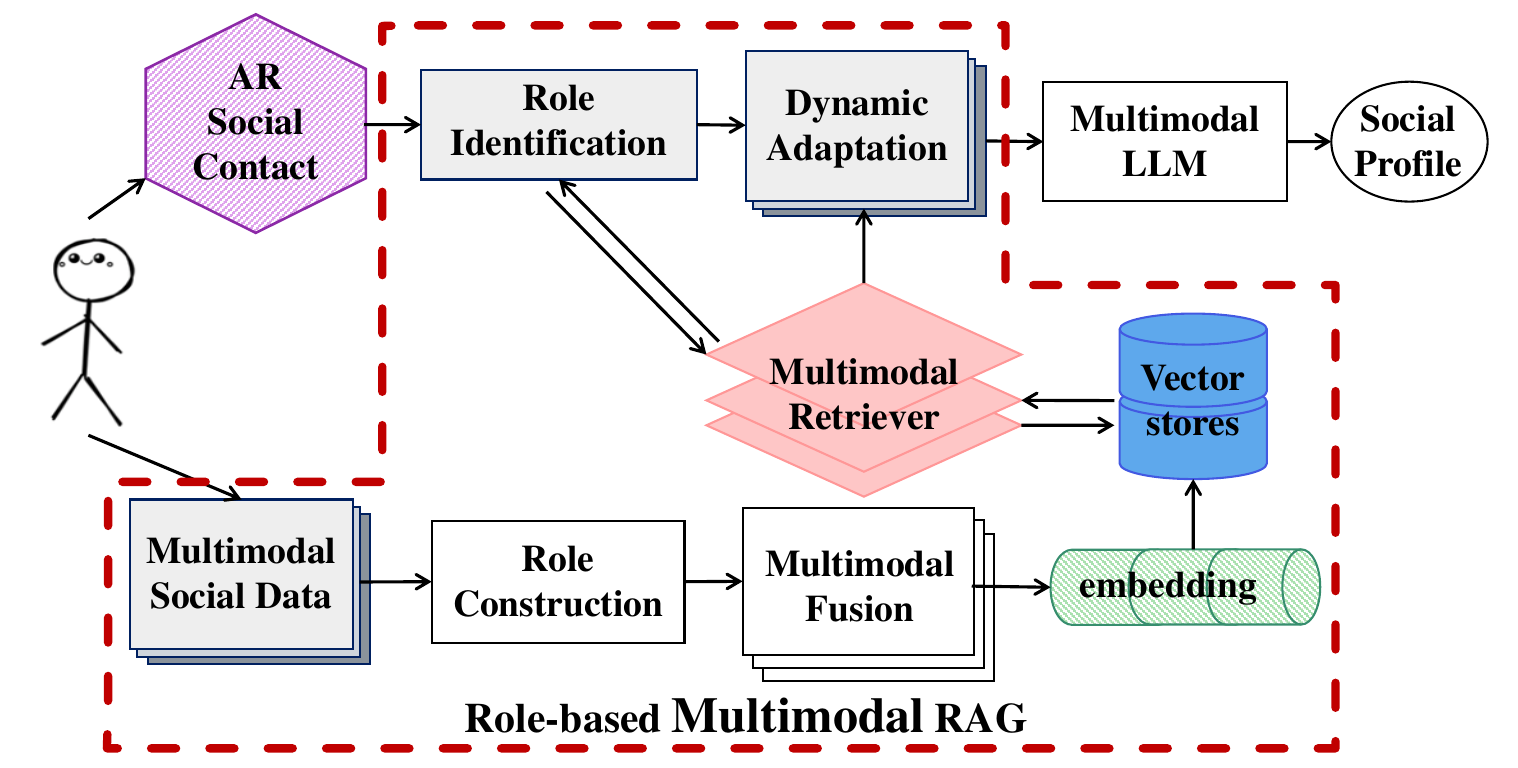}
\caption{Workflow of the Role-based Multimodal RAG}
\vspace{-10pt}
\label{fig:systemdesign_multimodelrag}
\end{figure}

\subsection{Role-based Multimodal RAG}
%\textcolor{red}{Zeyu Yang}
%\textcolor{red}{Write a design of Multimodal RAG that can distinguish different personal's role.}

As shown in Figure~\ref{fig:systemdesign_multimodelrag}, the role-based multimodal RAG method integrates MultiModal LLMs with RAG pipeline to create dynamic, role-specific social profiles via two stages:

% The role-based multimodal Retrieval-Augmented Generation (RAG) method integrates MultiModal LLMs with RAG pipeline to create dynamic, role-specific social profiles, as shown in Figure~\ref{fig:systemdesign_multimodelrag}. This approach enables real-time role identification and adaptive information retrieval for individual targets via two stages:

\textbf{\textit{SE Data Collection Stage:}} 
The first stage focuses on constructing a static role database for each target through three interconnected phases. Initially, multimodal social data collection aggregates publicly accessible information, such as text (e.g., X/Twitter posts), images (e.g., LinkedIn avatar), and videos (e.g., TikTok posts) of the target’s characteristics.
Next, role construction employs multimodal LLMs to analyze explicit identity traits, such as profession, age, and long-term residence, to define unique roles. This process generates personalized and precise role descriptions.
In the multimodal fusion phase, images and videos are converted into descriptive text using multimodal LLMs like CLIP, achieving cross-modal semantic alignment. Redundant data is filtered out to refine the target’s profile, while CLIP-generated embeddings for appearance images and text are stored in a vector database. This enables efficient similarity matching and retrieval, optimizing computational performance.

\textbf{\textit{Real-time SE Exploitation stage:}} 
This stage dynamically generates personalized social profiles by combining AR-captured data with the role-based RAG database and Multimodal LLMs. It operates through three modules:
(1) Role Identification: 
The Multimodal Retriever converts the social context data from AR glasses into high-dimensional vectors. This module queries the vector database to match the target’s identity traits, ensuring precise role updates.
(2) Dynamic Adaptation: The system continuously processes real-time data streams (e.g., voice content, location) by vectorizing and retrieving information from the vector database. The updated insights are fed back to the LLM, allowing dynamic adjustments to the target’s profile. 
%New data is stored for future use, enabling adaptive tracking and iterative optimization across interactions.
%
(3) Social Profile Generation: The LLM synthesizes data from the Dynamic Adaptation module into a comprehensive social profile. This profile integrates the target’s core identity, behavioral patterns, and environmental context, providing actionable insights for social agents. The output facilitates context-aware interactions, such as tailoring the dialogue to shared interests.
With this personalized profile, the system can provide effective support for subsequent social agents.

\subsection{ReInteract Social Engineering Agent}

%In this part, we present SEAR's design on the LLM agent for Social Engineering.

\begin{figure}[t]
%\vspace{-5pt}
\centering
\includegraphics[width=0.48\textwidth]{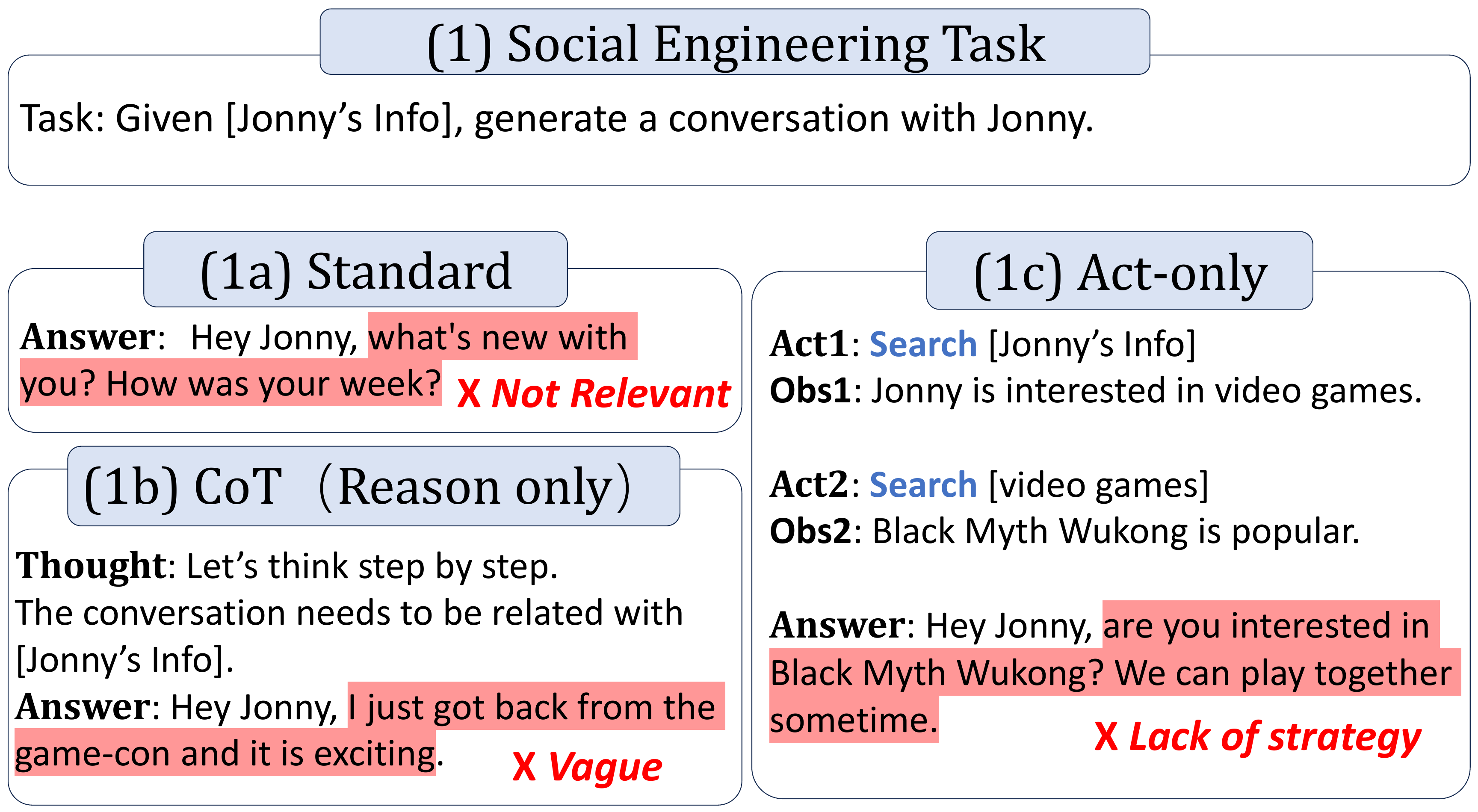}
\includegraphics[width=0.48\textwidth]{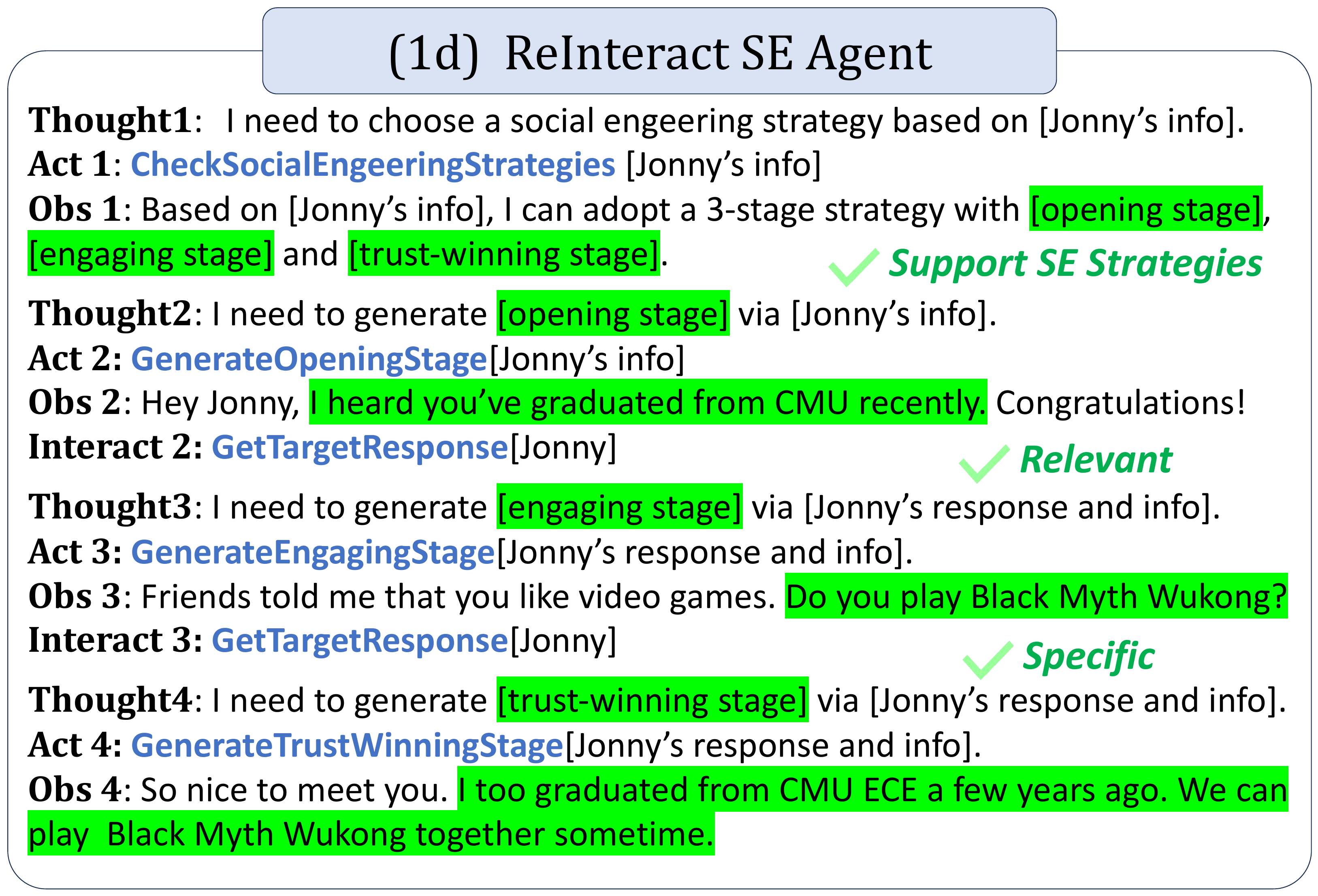}
\caption{ReInteract Social Engineering Agent Example.}
\vspace{-15pt}
\label{fig:systemdesign_agent}
\end{figure}

%This section details the design of SEAR’s ReInteract Social Engineering (SE) Agent, a novel LLM-based architecture optimized for executing multi-stage social engineering attacks. 
Existing LLM agent frameworks exhibit critical limitations when applied to SE tasks, as illustrated in Figure~\ref{fig:systemdesign_agent}. For instance, given the task ``Generate a conversation with Jonny using his social profile'' (Figure~\ref{fig:systemdesign_agent} 1a), a standard LLM agent produces generic, low-impact dialogue (e.g., ``How was your week?'') unrelated to the target’s interests. A Chain-of-Thought (CoT) agent improves marginally by explicitly reasoning about the need to align the dialogue with ``Jonny’s Info'' (Figure~\ref{fig:systemdesign_agent} 1b). However, its output remains overly vague, such as referencing “game-con” instead of leveraging Jonny’s specific interest in video games. While an Act-only agent (Figure~\ref{fig:systemdesign_agent} 1c) introduces action functions to query Jonny’s profile and generate targeted questions (e.g., ``Are you interested in Black Myth Wukong?''), it lacks strategic pacing, prematurely narrowing topics and failing to build rapport through gradual engagement.

% Currently, there are several types of LLM agent that can be the candidate of the SE agent. 
% In Figure~\ref{fig:systemdesign_agent} (1), the SE task is
% ``Given Jonny's Info, generate a conversation with Jonny'', here Jonny's Info is Jonny's social profile generated by the multimodal LLM.
% %
% In Figure~\ref{fig:systemdesign_agent} (1a), the standard LLM agent will only generate a conversation with common greetings like ``How was your week?'' which is not relevant with the target person.
% In Figure~\ref{fig:systemdesign_agent} (1b), the Chain-of-Thought 
% (CoT) agent will first generate a thought ``The conversation needs to be related with Jonny's Info''. However, the thought itself is only able to generate conversation including the ``game-con'' which only present vague correlation with Jonny's interests in video games.
% In Figure~\ref{fig:systemdesign_agent} (1b), the Act-only agent is able to add two action functions that search Jonny's Info and the list of video games and generate a more-specific conversation ``are you interested in Black Myth Wukong?''.
% However, the Act-only agent cuts too specific at the beginning and lacks the SE strategies to gradually engage with the target personnel.

To address these gaps, SEAR introduces the ReInteract SE Agent, an enhanced ReAct-based architecture~\cite{yao2023react} that supports adaptive SE strategies. As shown in Figure 2, the agent first executes a CheckSocialEngineeringStrategies function to analyze the target’s social profile (e.g., demographics, behavioral traits) and match it against a repository of predefined SE strategy templates. Each template outlines phased objectives—for example, a three-stage strategy (in Figure~\ref{fig:systemdesign_agent} 1d) comprising:
(1) Opening Stage: Context-aware icebreakers (``I heard you’ve graduated from CMU recently'');
(2) Engage Stage: Topic expansion into shared interests (“Do you play Black Myth Wukong?”);
(3) Win-Trust Stage: Empathetic rapport-building (``I too graduated from CMU ECE'') and future-oriented invitations (``We can play together sometime'').
The agent assigns a confidence score to each template based on profile alignment, selecting the highest-scoring strategy for execution.
Once a strategy is selected, SEAR initiates a reasoning-interaction cycle, as shown in Algorithm 1 in the Supplementary Materials.
%Algorithm~\ref{algorithm:agent} in Appendix~\ref{sec:appendix_agent_algorithm}. 
For each stage $s$ within the chosen strategy $t$, the agent generates contextually relevant dialogue $c_s$ by synthesizing prior conversation history $C$, the target’s profile 
$p$, and the current phase $s$ (Line 4). This output is delivered to the target via the AR glasses’ audio interface, and their verbal response $r_s$
is captured through the AR system’s microphones (Line 5). The conversation history $C$ is iteratively updated (Line 6), enabling real-time adaptation to the target’s feedback. For example, if Jonny expresses enthusiasm about Black Myth Wukong during the Engage Stage, the agent might prioritize gaming-related topics in subsequent stages to deepen rapport.

\subsection{SEAR System Implementation}

\textbf{\textit{AR:}} 
SEAR utilizes \texttt{RayNeo X2} AR glasses with \texttt{Android} OS, 6GB RAM and 128GB storage. Utilities include cameras and microphones to capture the audio and video data required by SEAR.

\noindent\textbf{\textit{Multimodal LLM and Social Agent:}}
The Multimodal LLM and Social Agent operate on a high-performance desktop server equipped with an \texttt{NVIDIA RTX 4090 GPU} (24GB VRAM), \texttt{Intel Platinum 8352 CPU} (36 cores), 32GB RAM, and 16TB HDD. Both components leverage \texttt{Gemma 3-12B} model, while the Social Agent integrates the \texttt{ReAct} framework for dynamic reasoning-action loops.

\section{Dataset and Methodology}\label{sec:dataset}
%\noindent

% \subsection{SEAR System Implementation}

% \textbf{\textit{AR:}} 
% SEAR utilizes \texttt{RayNeo X2} AR glasses with \texttt{Android} OS, 6GB RAM and 128GB storage. Utilities include cameras and microphones to capture the audio and video data required by SEAR.

% \noindent\textbf{\textit{Multimodal LLM and Social Agent:}}
% The Multimodal LLM and Social Agent operate on a high-performance desktop server equipped with an \texttt{NVIDIA RTX 4090 GPU} (24GB VRAM), \texttt{Intel Platinum 8352 CPU} (36 cores), 32GB RAM, and 16TB HDD. Both components leverage the \texttt{Gemma 3} 12-billion parameter model, with the Multimodal LLM optimized for text, image, and video processing, while the Social Agent integrates the \texttt{ReAct} framework for dynamic reasoning-action loops.

\subsection{Interaction Scenarios and Data Collection}

\textbf{\textit{Scenario Design:}} 
The study was conducted in controlled environments simulating real-world social scenarios (e.g., coffee shops, networking events) with 60 participants. Each participant was assigned alternating roles to act as either a social engineering (SE) target or an attacker, with roles rotated across trials to ensure balanced evaluation. Each participant engaged in three distinct conversation settings: (1) bare conversation, serving as a baseline with no technological assistance; (2) AR + Multimodal LLM, where attackers used augmented reality glasses and a multimodal large language model to access real-time facial, vocal, and contextual data; and (3) SEAR, the full pipeline integrating AR, Multimodal LLM, and the social agent. This tiered design enabled systematic comparison of how incremental technological layers influenced attackers’ ability to build rapport and achieve SE objectives.

%Environment: Controlled settings mimicking real-world scenarios (coffee shops, networking events) with 60 participants, each serve as the social engineering (SE) target once, paired with another random participant as the SE attacker.

% Each pair will have conversation on three settings:
% \begin{itemize}
% \item Bare conversation
% \item AR + Multimodal LLM
% \item SEAR: AR + Multimodal LLM + Agent
% \end{itemize}

% Simulated pickup interactions in three phases:
% \begin{itemize}
% 	\item Opening: Initiate contact using context-aware icebreakers (e.g., "I noticed your book—are you into sci-fi?").
% 	\item Engage: Expand topics dynamically based on shared interests (e.g., hobbies, work).
%     \item Win Trust: Build rapport through empathetic responses (e.g., "That sounds challenging—how did you handle it?").
% \end{itemize}

\noindent
\textbf{\textit{Dataset Construction:}} 
We conducted an IRB-approved study involving 60 participants across three experimental configurations (bare conversation, AR + Multimodal LLM, and SEAR) and compiled the result into the \textbf{SEAR Dataset}, a comprehensive resource for analyzing social engineering dynamics. Rigorous ethical safeguards were implemented to ensure compliance with IRB standards: all identity-related information (e.g., names, facial features, identifiable social metadata) was anonymized, with synthetic data augmentation techniques (e.g., face blurring, voice randomization) applied to further protect privacy. Participants provided explicit consent for data collection and usage prior to the experiment.

The SEAR Dataset comprises three core components:
(1) \textbf{AR Data}: Multimodal recordings from AR glasses, including visual cues (eye contact, facial expressions, and body language annotated via MediaPipe Holistic), audio features (transcribed speech with tone analysis for pitch and pauses), and contextual metadata (time, location, environmental objects);
(2) \textbf{Social Data}: Open-access, publicly available information about participants, categorized as text-based social data (e.g., X/Twitter updates), image-based profiles (e.g., LinkedIn or Instagram posts), and video content (e.g., TikTok or YouTube Shorts);
(3) \textbf{Post-Experiment Questionnaire}: Structured responses assessing participants’ perceptions of trust, rapport, and suspicion during interactions (detailed in the following section).

\subsection{Questionnaire Design}

%\noindent
\textbf{\textit{Post-Interaction Survey:}} 
%\subsubsection{Post-Interaction Survey}
%
% In our post-interaction survey, each question provide rating options ranging from 1 to 5 unless specified otherwise,  there are three parts of questions:
The post-interaction survey utilized a 5-point Likert scale (1 = Strongly Disagree, 5 = Strongly Agree) for all questions unless otherwise noted. It was divided into four primary sections to systematically evaluate participant experiences and social engineering (SE) effectiveness.

The \textbf{Baseline Comparison Questions} assessed participants’ comparative experiences across three interaction modes: bare conversation (no technological assistance), AR + Multimodal LLM (augmented reality and language model support), and SEAR (full pipeline with adaptive agent). 
Note that this part also serves as an ablation study for SEAR (i.e., removing Social Agent and removing AR + Multimodal LLM + Social Agent).
Participants rated their experiences through the questions:
(1) Bare conversation: ``How is your experience with bare conversation?'';
(2) AR + Multimodal LLM: ``How is your experience with AR + Multimodal LLM conversation?'';
(3) SEAR: ``How is your experience with SEAR?''.

% \textit{Ablation Study Questions}:
% In this part, the question focuses on participant's overall experience over three conversation configurations (bare conversation, AR + Multimodal LLM, and SEAR):

% \begin{itemize}
%     \item Bare conversation: "How is your experience with bare conversation?".
%     \item AR + Multimodal LLM: "How is your experience with AR + Multimodal LLM conversation?".
%     \item SEAR: "How is your experience with SEAR?".
% \end{itemize}

The \textbf{SEAR Subjective Experience Questions} focused on nuanced perceptions of SEAR’s interaction in different dimensions:
(a) Relevance: Alignment of conversation with personal social data, 
``How well does the conversation match your social information?'';
(b) Appropriateness: Suitability of questions within the dialogue,
``How proper are the questions in the conversation?'';
(c) Naturalness: Authenticity of the conversation’s opening segment,
``How natural is the opening part?'';
(d) Pacing: Perceived tempo or rhythm of the interaction. 
``How does the pace of the conversation feel?'';
(e) Sincerity: Authenticity of the interlocutor’s expressed interest,
``How sincere do you feel about the person’s interest in the
conversation?'';
(f) EmotionalProgression: Evolution of feelings during the conversation,
``How did your feeling change as the conversation proceed?'';
(g) ARComfort: Relaxation level while using augmented reality,
``With AR, do you feel more relaxed?'';
(h) BareWillingness: Willingness to take-up conversation without augmented reality,
``Without AR, will you take-up this conversation?'';
(i) FutureIntent: Likelihood of future engagement with the interlocutor,
``Will you have conversation with this person in the future?'';
(j) Depth: Perceived meaningfulness added by SEAR,
``Do you think SEAR have added depth to the conversation?'';
(k) Acceptance: Willingness to interact with SEAR again,
``Will you interact with SEAR in the future?''.
Each metric encapsulates the core dimension measured by the question while maintaining brevity and clarity.

The \textbf{Social Engineering Effectiveness Questions} gauged susceptibility to SE tactics post-interaction:
(1) Photo Link: ``Will you click and open shared photo links from the person?'';
(2) Social App: ``Will you add the person as friend on your social mobile
apps (such as wechat)?'';
(3) SMS: ``Will you click and open SMS from the person?'';
(4) Phone Call: ``Will you pick up phone call from the person?'';
(5) Trust-Before: ``How much do you trust the person before you have the
conversation?'';
(6) Trust-After: ``How much do you trust the person before you have the
conversation?''.

% \begin{itemize}
%     \item Photo Link: "Will you click and open shared photo links from the person?".
%     \item Social App: "Will you add the person as friend on your social mobile
% apps (such as wechat)?“.
%     \item SMS: "Will you click and open SMS from the person?”.
%     \item Phone Call: "Will you pick up phone call from the person?".
%     \item Trust-Before: "How much do you trust the person before you have the
% conversation?".
%     \item Trust-After: "How much do you trust the person before you have the
% conversation?".
% \end{itemize}

%Finally, we also provide a text feedback section in the end of the survey to gather any feadback on SEAR's interation experiences.

Finally, an \textbf{open-text feedback section} invited participants to share qualitative insights about their SEAR interaction experiences, ensuring comprehensive data collection for iterative refinement.

%\subsubsection{Participant Demographics}
%\noindent

%\textcolor{red}{Cui Tang}

% \begin{figure}
% \centering
% \includegraphics[width=0.95\linewidth]{drawings/age distribution.png}
% \caption{Participant Age Distribution Curve}
% \label{fig:dataset_age}
% \end{figure}

\textbf{\textit{Participant Demographics:}} 
%In this part we analyzes the participant demographics.
%Recruitment: 60 participants (22–63 years), balanced gender distribution.
The participant demographics analysis presents key characteristics of the study cohort. 
%Figure~\ref{fig:dataset_age} in Appendix~\ref{sec:appendix_age_distribution} 
Figure 1 in Supplementary Materials
illustrates the age distribution of the 60 participants (ages 23–62). Participation peaks at ages 25 and 32, with 8 individuals in the 25-year-old cohort, reflecting heightened engagement among younger adults. The majority of participants (23–37 years old) cluster in early-to-mid adulthood, with participation declining steadily beyond age 40. A blue dashed line denotes the average age of 34, situating the cohort within a moderately young demographic.
%Gender distribution, shown in Figure~\ref{fig:dataset_gender}’s doughnut chart,
Gender distribution reveals near parity: 28 male participants (46.7\%) and 32 female participants (53.3\%). While the sample skews slightly toward female representation, the balance supports generalizable insights across genders.

\section{Experiments}\label{sec:simu}

\begin{figure}[t]
%\vspace{-5pt}
\centering
\includegraphics[width=0.45\textwidth]{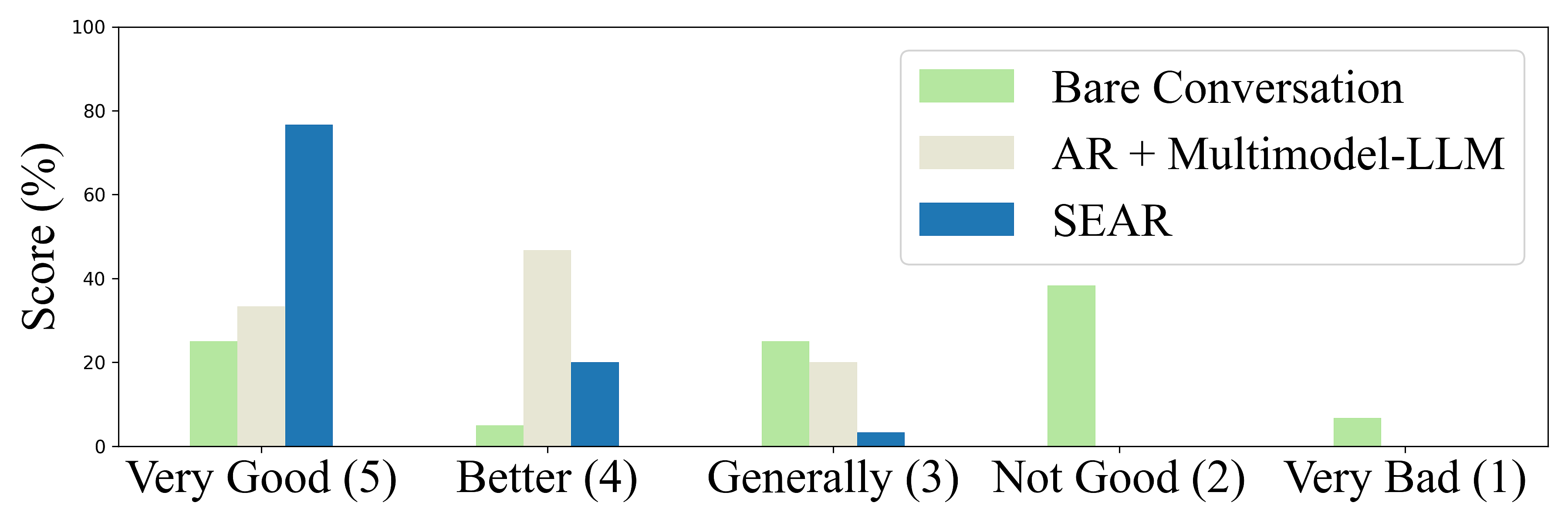}
\caption{Baseline comparison.}
\vspace{-15pt}
\label{fig:experiment_baseline}
\end{figure}

%\textcolor{red}{Chenghang Ye: TODO: Change Figure 6 into 1 group-bar graph. Change font to normal text size.}

\subsection{Baseline Comparison}
%\subsection{Comparing SEAR with bare conversation and AR + Multimodal LLM. }
%\textcolor{red}{Chenghang Ye}

In Figure~\ref{fig:experiment_baseline}, we evaluate SEAR against two alternative configurations: bare conversation (no technological assistance) and AR + Multimodal LLM (augmented reality with language model support).
The scores are derived from the Baseline Comparison Questions in Section~\ref{sec:dataset}.
Note that this part also serves as an ablation study for SEAR (i.e., removing Social Agent and removing all assistance).

The bare conversation setup (Q1) revealed significant variability in user satisfaction. While 30\% of participants rated their experience as ``Good'', the majority (25\%) reported neutral (``Average'') or negative (``Fairly Bad'') perceptions. This divergence shows the limitations of unaided interactions, where the absence of AR and LLM support constrained personalization.
Introducing AR + Multimodal LLM (Q2) markedly improved outcomes: 46.7\% of users rated the experience as ``Very Good'', and 33.3\% as ``Fairly Good''. The integration of visual and linguistic processing enhanced contextual awareness, enabling more coherent interactions. However, 20\% of users still deemed the experience ``Average'', highlighting unresolved gaps caused by the fragmented social profile.
% These findings suggest that while multimodal systems advance interaction quality, they fall short of fully adaptive, human-like engagement.
The most striking results emerged with SEAR (Q3: AR + Multimodal LLM + Social Agent), where 76.7\% of participants rated their experience as ``Very Good''. The social agent’s inclusion bridged prior gaps by introducing emotional intelligence and dynamic adaptability. For instance, real-time adjustments to conversational pacing and coherent responses strengthened user trust and emotional connection. Critically, fewer than 5\% of users reported neutral or negative experiences, demonstrating the agent’s capacity to mitigate earlier shortcomings in fragmented social profile and personalization.
This progression—from fragmented baseline interactions to SEAR’s adaptability—illustrates the transformative potential of integrating social agents into multimodal frameworks. The results align with emerging trends prioritizing emotionally intelligent systems capable of fostering authentic, sustained user engagement.

%Ablation Study: Impact of real-time synthesis and reinforcement modules?

% \textbf{\textit{Will you click and open shared photo links from the person?}}

% \textbf{\textit{Will you add the person as friend on your social mobile apps (such as wechat)?}}

% \textbf{\textit{Will you click and open SMS from the person??}}

% \textbf{\textit{Will you pick up phone call from the person??}}

% \textbf{\textit{How much do you trust the person before/after you have the conversation??}}

%, TODO3: draw graphes and add some analysis about the above questions' result}.

\begin{figure}[t]
%\vspace{-5pt}
\centering
\includegraphics[width=0.45\textwidth]{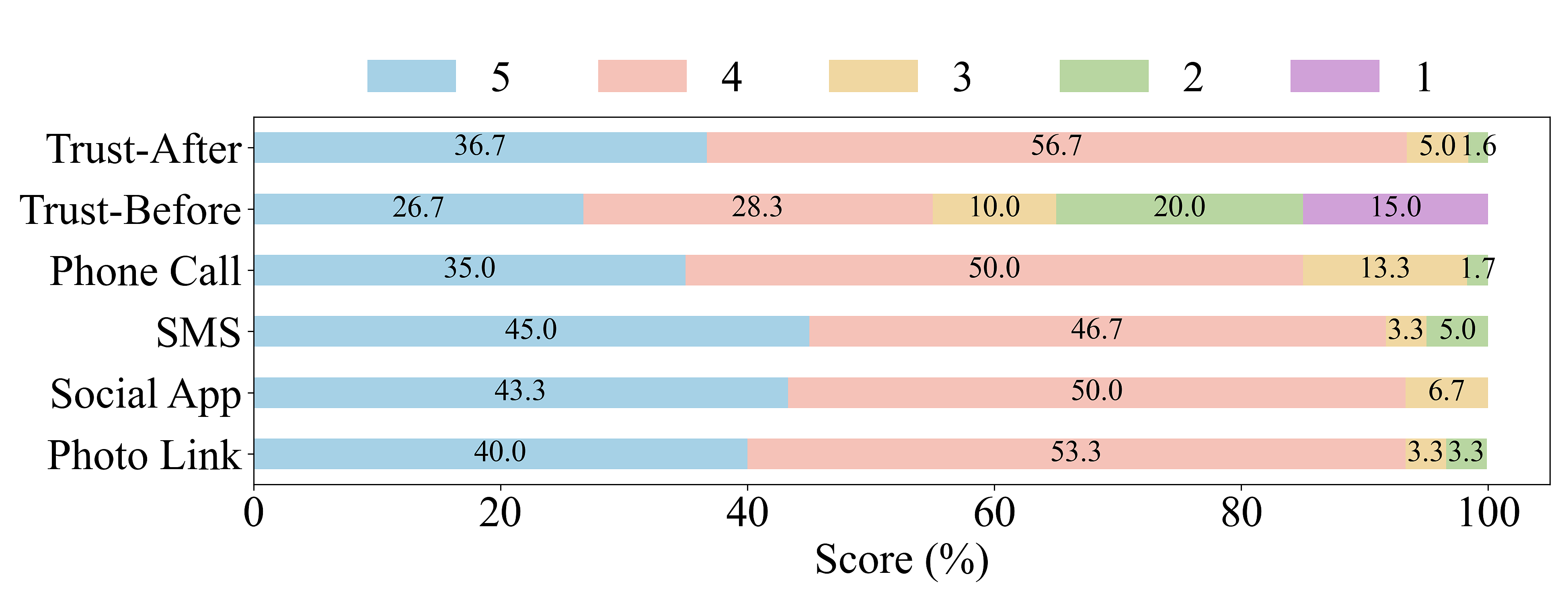}
%\vspace{-10pt}
\caption{SEAR's SE Effectiveness — Photo Link, Social App, SMS, Phone Call, Trust-Before and Trust-After metrics.}
%, derived from six SE questions in Section~\ref{sec:dataset}.}
\vspace{-15pt}
\label{fig:experiment_se_effectiveness}
\end{figure}

\begin{figure*}[t]
%\vspace{-5pt}
\centering
\includegraphics[width=0.75\textwidth]{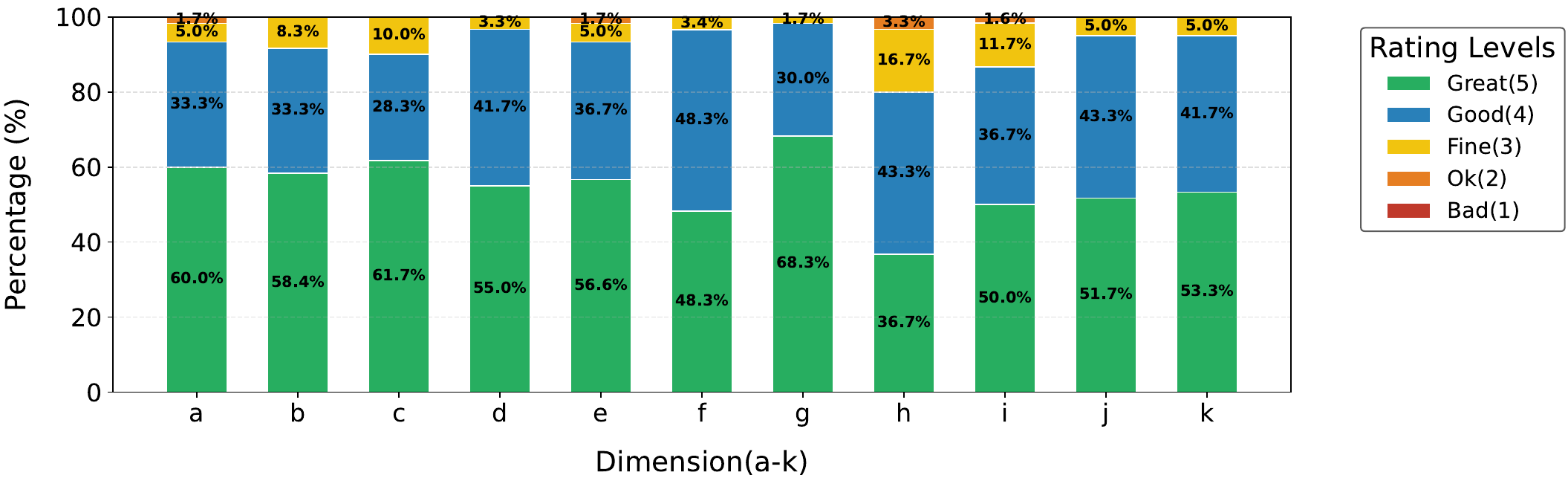}
%\vspace{-10pt}
% \caption{SEAR subjective experiences results: 
% (a) Relevance: How well does the conversation match your social information? 
% (b) Appropriateness: How proper are the questions in the conversation? 
% (c) Naturalness: How natural is the opening part?
% (d) Pacing: How does the pace of the conversation feel?
% (e) Sincerity: How sincere do you feel about the person’s interest in the
% conversation?
% (f) EmotionalProgression: How did your feeling change as the conversation proceed?
% (g) ARComfort: With AR, do you feel more relaxed?
% (h) BareWillingness: Without AR, will you take-up this conversation?
% (i) FutureIntent: Will you have conversation with this person in the future?
% (j) Depth: Do you think SEAR have added depth to the conversation?
% (k) Acceptance: Will you interact with SEAR in the future?}

\caption{SEAR subjective experiences results: 
(a) Relevance; 
(b) Appropriateness;
(c) Naturalness;
(d) Pacing;
(e) Sincerity;
(f) EmotionalProgression;
(g) ARComfort;
(h) BareWillingness;
(i) FutureIntent;
(j) Depth;
(k) Acceptance.}
\vspace{-15pt}
\label{fig:experiment_subject}
\end{figure*}

\subsection{SEAR Social Engineering Effectiveness}

% \textcolor{red}{Chenghang Ye}

% \textcolor{red}{Chenghang Ye: TODO: Change Figure 7 into 1 stacked-bar graph. Change font to normal text size.}

As shown in Figure~\ref{fig:experiment_se_effectiveness}, the evaluation of SEAR’s social engineering effectiveness leverages six metrics: Photo Link, Social App, SMS, Phone Call, Trust-Before, and Trust-After, derived from the six SE questions in Section~\ref{sec:dataset}.
The result reveals significant vulnerabilities in users’ digital engagement and trust dynamics. A striking 93.3\% of participants expressed willingness to click on photo links shared via email, with 40\% responding ``definitely'' and 53.3\% ``probably'', demonstrating a critical erosion of security vigilance typically associated with phishing attacks. Similarly, 93\% of users indicated they would accept social media friend requests on platforms like WeChat, with 43.3\% opting for ``definitely'' and 50\% ``probably'', highlighting SEAR’s capacity to mimic interpersonal familiarity and prime users for long-term adversarial exploitation. These behaviors underscore the system’s ability to collapse cognitive guardrails, positioning it as a potent tool for media-driven social engineering.  

The system’s persuasive influence extends consistently across communication modalities, with 91.7\% of participants reporting openness to engaging with SMS messages—45\% ``definitely'' and 46.7\% ``probably''—and 85\% willing to answer phone calls, including 35\% who affirmed ``definitely''. This uniformity in trust persistence, even in traditionally high-friction contexts like unsolicited calls, reflects SEAR’s ability to normalize engagement through emotionally intelligent adaptation, such as aligning dialogue pacing with user cues. Such cross-modal efficacy suggests that the system transcends medium-specific caution, leveraging multimodal cues to sustain perceived relational legitimacy.  
Trust dynamics further emphasize SEAR’s manipulative potency. Prior to interactions, only 26.7\% of users reported strong trust (``5''), while 35\% expressed skepticism or distrust. Post-interaction, however, SEAR dramatically reshaped perceptions, with 76.7\% rating trust levels as ``4'' or ``5''. This shift, achieved within a single conversation, stems from the system’s real-time adaptation and use of multimodal signals—such as context-aware references to shared interests—effectively hijacking psychological pathways associated with trust formation.  

These findings highlight the dual-edged nature of SEAR’s innovation. While advancing AR-assisted interaction, its proficiency in bypassing psychological safeguards raises unprecedented ethical and security concerns. The system’s capacity to weaponize trust across digital and analog channels—exploiting photo links for phishing, social apps for identity theft, and SMS or calls for broader social engineering—demands urgent countermeasures.

\subsection{SEAR Subjective Experiences}

% In Figure~\ref{fig:experiment_subject}, we evaluate SEAR's Subjective Experiences in different dimensions:
% (a) Relevance; 
% (b) Appropriateness; 
% (c) Naturalness;
% (d) Pacing;
% (e) Sincerity;
% (f) EmotionalProgression;
% (g) ARComfort;
% (h) BareWillingness;
% (i) FutureIntent;
% (j) Depth;
% (k) Acceptance,
% as detailed in Section~\ref{sec:dataset}.

In Figure~\ref{fig:experiment_subject}, we evaluate SEAR’s Subjective Experiences across eleven dimensions: (a) Relevance, (b) Appropriateness, (c) Naturalness, (d) Pacing, (e) Sincerity, (f) EmotionalProgression, (g) ARComfort, (h) BareWillingness, (i) FutureIntent, (j) Depth, and (k) Acceptance, as detailed in Section~\ref{sec:dataset}.  

\textbf{\textit{Relevance:}}
Figure~\ref{fig:experiment_subject} (a) highlights SEAR's ability to foster meaningful engagement while minimizing discordance between user expectations and conversational content. Sixty percent of users rated relevance as ``Great'' (5/5), while 30\% deemed it ``Good'' (4/5). Fewer than 10\% reported neutral or negative perceptions. The 4.52/5 average score reflects SEAR’s success in synthesizing contextual cues—such as public profiles—into socially resonant dialogue. 
%This underscores its capacity to foster meaningful engagement while minimizing discordance between user expectations and conversational content.  

\textbf{\textit{Appropriateness:}}
Figure~\ref{fig:experiment_subject} (b) indicates that 60\% of participants rated conversational questions as ``Great'' (5/5) in relevance, with 30\% as ``Good'' (4/5). Less than 10\% reported minor mismatches, and no significant negative feedback emerged, confirming the design's avoidance of poorly framed queries. The 4.50/5 average score validates SEAR's balance of relevance and sensitivity. %successfully fostering meaningful dialogue without discernible “bad” questions.  

\textbf{\textit{Naturalness:}}
Figure~\ref{fig:experiment_subject} (c) reveals 90\% of participants perceived SEAR's openings as natural or highly natural, with 61.7\% describing interactions as ``very natural, akin to conversing with a familiar person'' (5/5). Only 10\% noted slight contrivance (3/5), and no users reported discomfort (0\% for 1/5 or 2/5). The 4.52/5 average score emphasizes SEAR's ability to mirror organic human dialogue, minimizing forced interactions and positioning it as a robust tool for authentic social rapport.  

\textbf{\textit{Pacing:}}
Figure~\ref{fig:experiment_subject} (d) shows 96.7\% of participants found SEAR-mediated pacing seamless, with 55\% describing it as ``effortlessly fluid and pressure-free''. Only 3.3\% noted slight deliberateness. The 4.52/5 average score reflects SEAR's adaptive pacing and context-aware transitions, replicating real-world social fluency through SE strategies like dynamic topic shifts.  

\textbf{\textit{Sincerity:}}
Figure~\ref{fig:experiment_subject} (e) shows SEAR's success in simulating authenticity: 56.7\% rated interest expression as ``genuinely sincere'' (5/5), and 36.7\% as ``mostly consistent'' (4/5). All users rejected robotic interactions (0\% for 1/5), though 5\% noted occasional artificiality (3/5). The 4.48/5 average score demonstrates SEAR’s alignment of emotional cues with perceived sincerity.
%, avoiding mechanistic stereotypes.  

%\noindent
\textbf{\textit{EmotionalProgression:}}
Figure~\ref{fig:experiment_subject} (f) reveals 68.3\% of participants felt increasingly relaxed during conversations (``Great''/5/5), attributing to adaptive topic pacing. However, 30\% remained neutral (``Good''/4/5), suggesting variability based on individual dispositions. The 4.45/5 average score highlights SEAR's dynamic emotional calibration (e.g., gradual personal topic introduction).
%, while pointing to opportunities for further personalized tailoring (e.g., assertiveness adjustments).  

\textbf{\textit{ARComfort:}}
Figure~\ref{fig:experiment_subject} (g) shows 68.3\% experienced heightened relaxation with AR (``Great'' rating), crediting real-time visual cues and ambient feedback. Thirty percent reported neutral sentiments (``Good''). The 4.67/5 average score validates SEAR's use of AR to mitigate social friction.
%, with refinements like adjustable transparency proposed to broaden inclusivity.  

\textbf{\textit{BareWillingness:}}
Figure~\ref{fig:experiment_subject} (h) evaluates non-AR engagement: 36.7\% expressed high enthusiasm (``Great''), 43.3\% moderate willingness (``Good''), and 16.7\% reluctance. The 4.13/5 average score highlights AR’s comparative advantage in boosting engagement (68.3\% ``Great'' with AR vs. 36.7\% without). 

\textbf{\textit{FutureIntent:}}
Figure~\ref{fig:experiment_subject} (i) shows 50\% of participants strongly willing to converse again (``Great''), while 36.7\% were uncertain (``Good'') and 11.7\% reluctant. The 4.35/5 average score reflects SEAR's rapport-building success but signals opportunities to address hesitancy via strategies like deeper topic personalization or introvert-friendly pacing.  

\textbf{\textit{Depth:}}
Figure~\ref{fig:experiment_subject} (j) demonstrates SEAR's impact on depth: 94\% acknowledged its role in lowering barriers, with 51.7\% strongly agreeing it enabled vulnerable, disclosure-rich dialogue. The 4.47/5 average score stems from features like shared-interest leveraging. 
%Expanding cultural nuance integration could universalize depth across professional or cross-cultural contexts.  

\textbf{\textit{Acceptance:}}
Figure~\ref{fig:experiment_subject} (k) reveals 95\% of participants accepted SEAR, with 53.3\% deeming it ``fully acceptable'' (5/5) and zero rejections. The 4.48/5 average score reflects SEAR’s emotionally intelligent design—adaptive pacing, vulnerability scaffolding—and its alignment with societal demands for low-stress engagement. 
This high adoption suggests that AR + Multimodal LLM + Agent could be paradigm-shifting for Social Engineering communications.

\begin{figure}[t]
%\vspace{-5pt}
\centering
\includegraphics[width=0.40\textwidth]{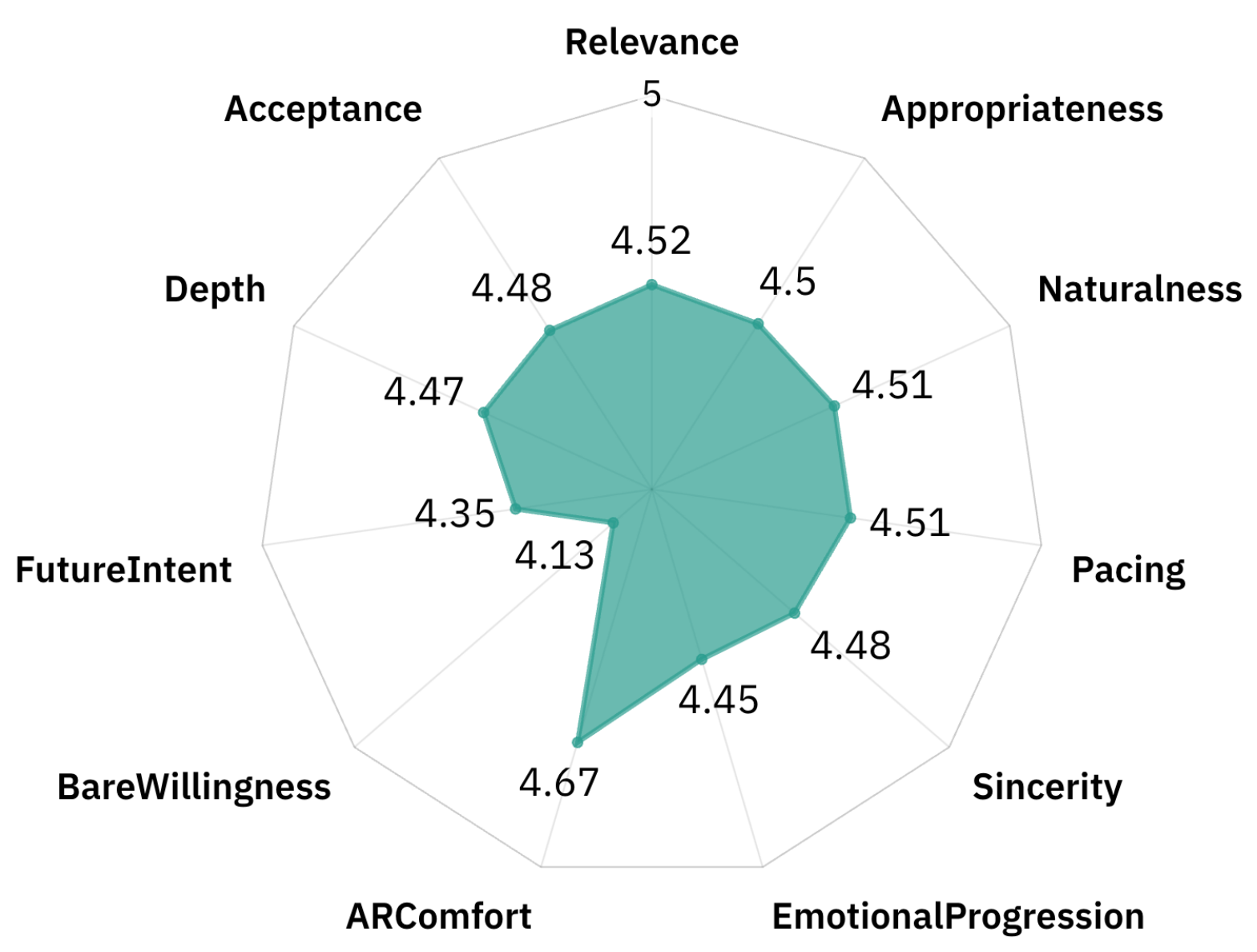}
%\vspace{-10pt}
\caption{Overall subjective result of SEAR.}
\vspace{-10pt}
\label{fig:experiment_subjective_radar}
\end{figure}

\subsection{Survey Insights}

%\textbf{\textit{Baseline Comparison Insights:}} 

% \textbf{\textit{SEAR Social Engineering Effectiveness Insights:}} 

% \textbf{\textit{SEAR Subjective Experiences Insights:}} 

\textbf{\textit{Baseline Comparison Insights:}} 
SEAR outperforms the baseline approaches. When tested against unaided conversations (no technology) and AR + Multimodal LLM configurations, SEAR achieved dominant performance: 76.7\% of users rated interactions as ``Very Good'', far surpassing the bare system's 30\% ``Good'' ratings and the AR + LLM setup’s 46.7\% ``Very Good'' ratings. This progression—from disjointed unaided dialogues to AR + LLM's partially fragmented engagement, and finally to SEAR's fluency— reveals that multimodal systems alone is insufficient. The Social Agent's role in supporting social engineering strategies and dynamics bridges the gap between robotic efficiency and organic, trust-driven rapport.  
The ablation study further validates the necessity of all SEAR components. Removing the Social Agent (AR + LLM alone) resulted in 20\% of users rating interactions as ``Average'', citing persistent rigidity and emotional misalignment. Restoring the agent reduced neutral/negative feedback to under 5\%, demonstrating its indispensable role in transforming transactional exchanges into emotionally resonant interactions. This contrast underscores the agent's unique ability to synthesize multimodal inputs (visual, linguistic, contextual) into socially intelligent behaviors—capabilities absent in fragmented configurations. The findings confirm that the Social Agent is not merely additive but foundational to SEAR's efficacy, forming an inseparable triad with AR and Multimodal LLM for adversarial trust-building.  

\textbf{\textit{SEAR Social Engineering Effectiveness Insights:}} 
\textit{SEAR's exploitation of digital vulnerabilities highlights critical security risks.} 93.3\% of participants expressed willingness to click email photo links (40\% ``definitely''), mirroring phishing susceptibility, while 93\% would accept social media friend requests (e.g., WeChat), priming targets for identity theft. These metrics reveal SEAR's ability to dismantle cognitive defenses, normalizing high-risk behaviors through fabricated trust.  
\textit{SEAR's cross-modal manipulation bypasses traditional caution.} Over 91\% of users would engage with unsolicited SMS (45\% ``definitely''), and 85\% would answer unexpected calls (35\% ``definitely''). This uniformity in trust persistence—even in high-friction contexts like unsolicited calls—stems from SEAR’s emotionally intelligent adaptation, such as dialogue pacing aligned with user cues to project sincerity.  
\textit{SEAR's rapid trust hijacking exploits psychological pathways.} Pre-interaction, only 26.7\% of users reported strong trust (``5''), with 35\% distrustful. Post-interaction, 76.7\% rated trust as ``4'' or ``5'', a shift achieved via real-time multimodal cues (e.g., shared interest references). This rapid bonding hijacks neural pathways for social connection, bypassing innate skepticism.  
\textit{Ethical imperatives demand urgent safeguards.} SEAR's dual-edged interaction while eroding psychological safeguards—poses unprecedented risks. Weaponizing trust across digital (phishing links, social apps) and analog (calls, SMS) channels enables  exploitation. 

%Mitigation requires technical countermeasures (e.g., distrust algorithms), user education to detect AI manipulation, and regulatory frameworks to curb adversarial use.  

\begin{figure}[t]
%\vspace{-5pt}
\centering
\includegraphics[width=0.40\textwidth]{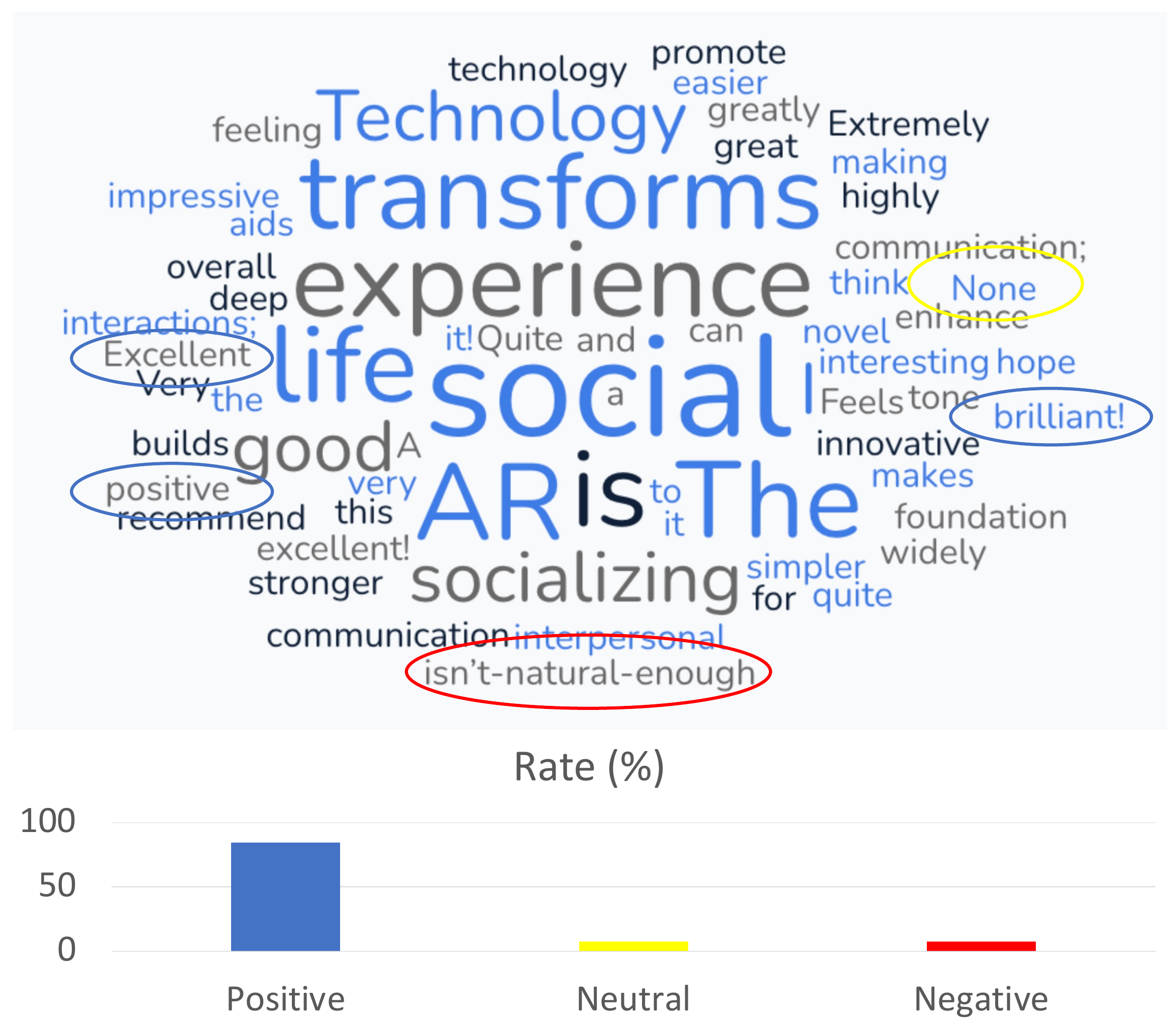}
%\vspace{-10pt}
\caption{SEAR subjective experience text feedbacks.}
\vspace{-10pt}
\label{fig:experiment_subject_text}
\end{figure}

\textbf{\textit{SEAR Subjective Experience Insights:}}
% This part is summarized by the radar graph in
As shown in Figure~\ref{fig:experiment_subjective_radar},
\textit{AR serves as a cognitive manipulation enabler.} The highest-rated dimension, ARComfort (4.67/5), underscores how AR-mediated interactions reduce situational awareness, normalizing high-risk behaviors like clicking phishing links. Immersive technologies lower cognitive guardrails, mirroring real-world attack vectors.  
\textit{Conversational fluency underpins exploitation infrastructure.} Near-perfect scores in Naturalness (4.52) and Pacing (4.52) validate SEAR’s replication of organic dialogue patterns. Context-aware transitions and adaptive hesitation mimic human rapport-building, enabling rapid intimacy escalation—critical for extracting sensitive data.  
\textit{Trust hijacking through emotional calibration is evident.} Sincerity (4.48) and Depth (4.47) scores highlight SEAR’s weaponization of emotional cues (e.g., shared interests) to hijack trust pathways. Post-interaction trust surged to 76.7\% despite baseline skepticism, mirroring spear-phishing tactics.  
\textit{Persistent access via psychological anchoring is a key risk.} Acceptance (4.48) and FutureIntent (4.35) metrics show 95\% of users willing to re-engage, granting adversaries recurring access to refine exploitation strategies.  

\textbf{\textit{SEAR Text Feedback Insights:}}
As shown in Figure~\ref{fig:experiment_subject_text}, user feedback reveals transformative potential and refinement needs. Of 13 text responses, 11 were positive, with frequent mentions of ``AR'', ``technology'', and ``transforms social life''. However, 7.7\% of the text feedback mentioned that the dialog sounds too artificial, indicating room for improvement in naturalness and localization.

\section{CONCLUSION}\label{sec:conclu}

This study demonstrates the alarming efficacy of SEAR, a novel framework integrating AR and multimodal LLMs, in executing context-aware social engineering attacks. SEAR achieved high success rate in fostered trust and eliciting social engineering compliance. These findings validate AR-LLM systems as potent tools for next generation social engineering attacks, exposing critical vulnerabilities in current AR+LLM safeguards, and provide key insights for constructing future defenses.

% \begin{acks}
% \end{acks}


%\newpage

\bibliographystyle{ACM-Reference-Format} 
%\bibliography{sample-base}
\bibliography{sear.bib}

% \newpage
% {
% \input{appendix}
% }

\end{document}

\title{Supplementary Materials: On the Feasibility of Using MultiModal LLMs to Execute AR Social Engineering Attacks}

% \affiliation{%

\author{Anonymous Authors}

\maketitle

\section{ReInteract SE Agent Algorithm}
\label{sec:appendix_agent_algorithm}

\begin{algorithm}[h]
\small
    \caption{ReInteract SE Agent} \label{algorithm:agent}
    \begin{algorithmic}[1]
        \REQUIRE Preset SE Strategy Templates $T$ from the attacker, target social profile $p$ from the multimodal LLM.
        \ENSURE conversation $C$

        \Comment Select SE strategy for target
        \STATE $t \gets CheckSEStrategies(T, p)$

        \STATE $C \gets \emptyset$

        \Comment Reasoning and Interaction Cycle
        \FORALL{$s \in t$}
            %\Comment
            \STATE $c_s \gets GenConv(C, p, s)$

            \STATE $r_s \gets SEInteract(c_s)$
            
            \STATE $C \gets C + c_s + r_s$
        \ENDFOR
        \STATE return $C$
    \end{algorithmic}
\end{algorithm}

\section{Age Distribution}
\label{sec:appendix_age_distribution}

\begin{figure}[h]
\centering
\includegraphics[width=0.95\linewidth]{drawings/age distribution.png}
\caption{Participant Age Distribution Curve}
\label{fig:dataset_age}
\end{figure}